\documentclass[a4paper]{jpconf}
\usepackage{graphicx}
\usepackage{wasysym}

\usepackage[colorlinks=true,linkcolor=blue]{hyperref}

\begin{document}
\title{A highly granular SiPM-on-tile calorimeter prototype}

\author{Felix Sefkow$^1$ and Frank Simon$^2${\\ \small On behalf of the CALICE Collaboration}}

\address{$^1$DESY, Hamburg, Germany\\ $^2$Max-Planck-Institut f\"ur Physik, Munich, Germany}

\ead{felix.sefkow@desy.de}

\begin{abstract}
The Analogue Hadron Calorimeter (AHCAL) developed by the CALICE collaboration is a scalable engineering prototype for a Linear Collider detector. It is a sampling calorimeter of steel absorber plates and plastic scintillator tiles read out by silicon photomultipliers (SiPMs) as active material (SiPM-on-tile). 
The front-end chips are integrated into the active layers of the calorimeter and are designed for minimizing power consumption by rapidly cycling the power according to the beam structure of a linear accelerator. 
38 layers of the sampling structure are equipped with cassettes containing 576 single channels each, arranged on readout boards and grouped according to the 36 channel readout chips. The prototype  has been assembled using techniques suitable for mass production, such as injection-moulding and semi-automatic wrapping of scintillator tiles, assembly of scintillators on electronics using pick-and-place machines and mass testing of detector elements.
The calorimeter was commissioned at DESY and was taking data at the CERN SPS at the time of the conference.  The contribution  discusses the construction, commissioning and first test beam results of the CALICE AHCAL engineering prototype.

\end{abstract}

\section {Introduction}

The physics at future high-energy lepton colliders, with its requirement for a  jet energy reconstruction with unprecedented precision, is one of the primary motivations for the development of highly granular calorimeters by the CALICE collaboration. The detector concepts for the International Linear Collider (ILC) and the Compact Linear Collider (CLIC) rely on Particle Flow Algorithms (PFA)~\cite{Brient:2002gh, pfaMorgunov, Thomson:2009rp}, which are capable of achieving the required resolution. This event reconstruction technique requires highly granular calorimeters to deliver optimal performance. 

One of the technologies developed within CALICE is the Analogue Hadron Calorimeter AHCAL. It is based on active elements consisting of 3$\times$3~cm$^2$ plastic scintillator tiles individually read out by silicon photomultipliers (SiPMs) in a steel absorber structure with approximately 20 mm of absorber material between each active layer. A cubic-metre sized "physics prototype"~\cite{Adloff:2010hb}, which has been extensively tested in particle beams at DESY, CERN and Fermilab, has demonstrated the capabilities of this technology, achieving competitive single hadron energy resolution \cite{Adloff:2012gv} and the two-particle separation required for good PFA performance \cite{Adloff:2011ha}. The prototype was also successfully tested with tungsten absorbers~\cite{Adloff:2013jqa, Blaising:2015nla}. 
An overview of the results can be found in \cite{Sefkow:2015hna}.

\section{The CALICE SiPM-on-Tile Hadron Calorimeter Engineering Prototype}
With the establishment of the principal viability of the AHCAL technology, the focus has shifted from the study of the physical performance characteristics of such a detector to the demonstration of the feasibility of the detector concept while satisfying the spatial constraints and scalability requirements of collider experiments~\cite{Behnke:2013lya, Linssen:2012hp}. Such a concept must be based on a scintillator tile design well-suited for mass production and automatic assembly, originally proposed in~\cite{Blazey:2009zz} and subsequently varied and optimised in further studies \cite{Simon:2010hf, Liu:2015cpe}. 

The new AHCAL prototype consists of a non-magnetic stainless steel absorber structure with 38 active layers and has 21888 channels. 
The basic unit of the active elements is the HCAL Base Unit HBU~\cite{6829522}, with a size of 36 $\times$ 36 cm$^2$, holding 144 SiPMs controlled by four SPIROC2E ASICs \cite{Bouchel:2011zz}.  A key element of the electronics is the capability for power-pulsed operation to reduce the power consumption and eliminate the need for active cooling, making use of the low duty cycle in the linear collider beam time structure. In addition to dual-gain energy measurement, the electronics also provides a cell-by-cell auto trigger and time stamping on the few ns level in test beam operations. In operating conditions with shorter data-taking windows closer to the bunch train structure of linear colliders, sub-ns time resolution is expected. 

\begin{figure}[hb]
\centering
\includegraphics[width=0.50\textwidth]{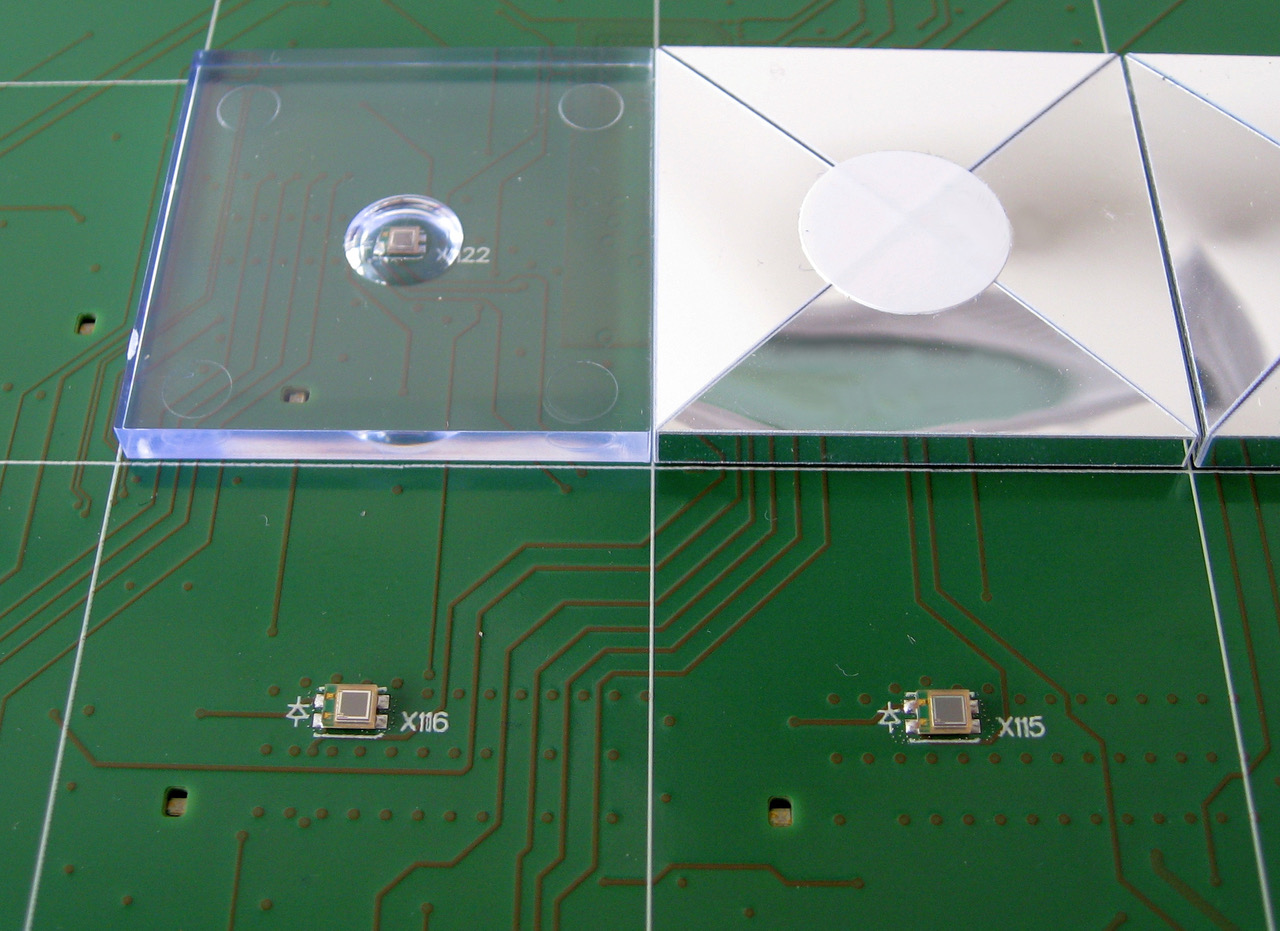}
\caption{CALICE AHCAL scintillator tiles with central dimple, wrapped and unwrapped, mounted on an HBU with SiPMs.}
\label{fig:Tile}
\end{figure}

The prototype uses Hamamatsu MPPC S13360-1325PE photon sensors and injection-moulded polystyrene scintillator tiles with a central dimple \cite{Liu:2015cpe} for optimal light collection, as shown in Figure \ref{fig:Tile}. 

The SiPMs were delivered in lots of 600 pieces with a uniform break-down voltage within $\pm 100$~mV. Spot-samples of all SiPM lots, and each one of the ASICs, had undergone semi-automatic testing procedures before soldering the HBUs \cite{Munwes:2634923}. The gain of the SiPMs was found to be uniform within~2.4\% when operated at a common over-voltage.
Without any further surface treatment, the scintillator tiles are wrapped in laser-cut reflective foil by a robotic procedure and mounted on the HBUs using a pick-and-place machine, after glue dispensing with a screen printer.   

\begin{figure}
\centering
\includegraphics[width=0.7\textwidth]{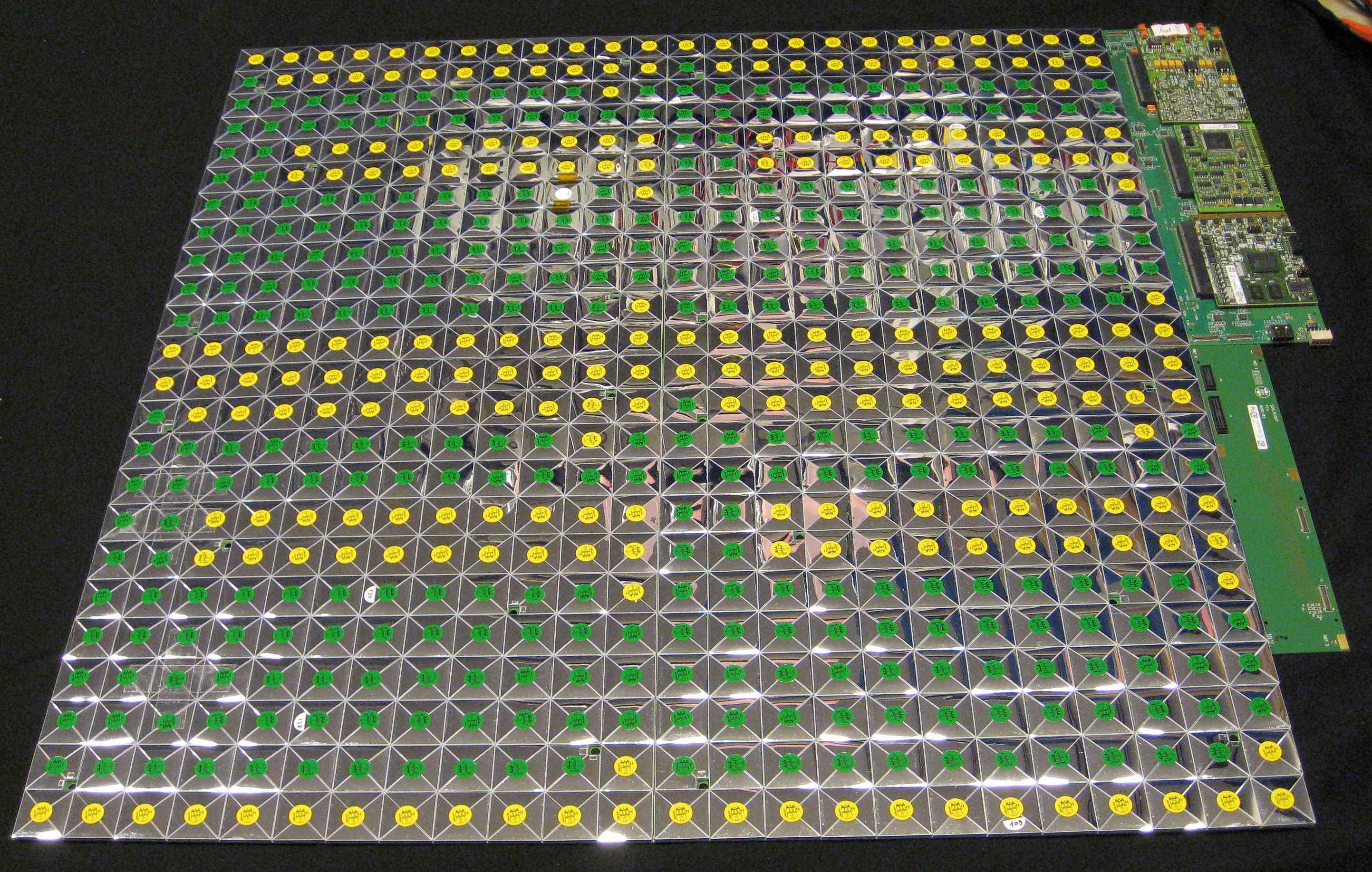}
\caption{Active layer side with wrapped tiles on 4 HBUs, with interfaces for DAQ, LEDs and power.}
\label{fig:ActiveLayer}
\end{figure}

The HBUs have been integrated into cassettes with interfaces for DAQ \cite{Kvasnicka:2017bpx}, LED pulsing and power distribution, which provide active compensation of temperature variations by automatic adjustments of the common bias voltage of the photon sensors in each layer. Figure~\ref{fig:ActiveLayer} shows the top side of one active layer, with the scintillator tiles visible. All layers have been  calibrated in the DESY test beam, and 99.96\% of the total 21888 channels are working.

\begin{figure}
\centering
\includegraphics[width=0.7\textwidth]{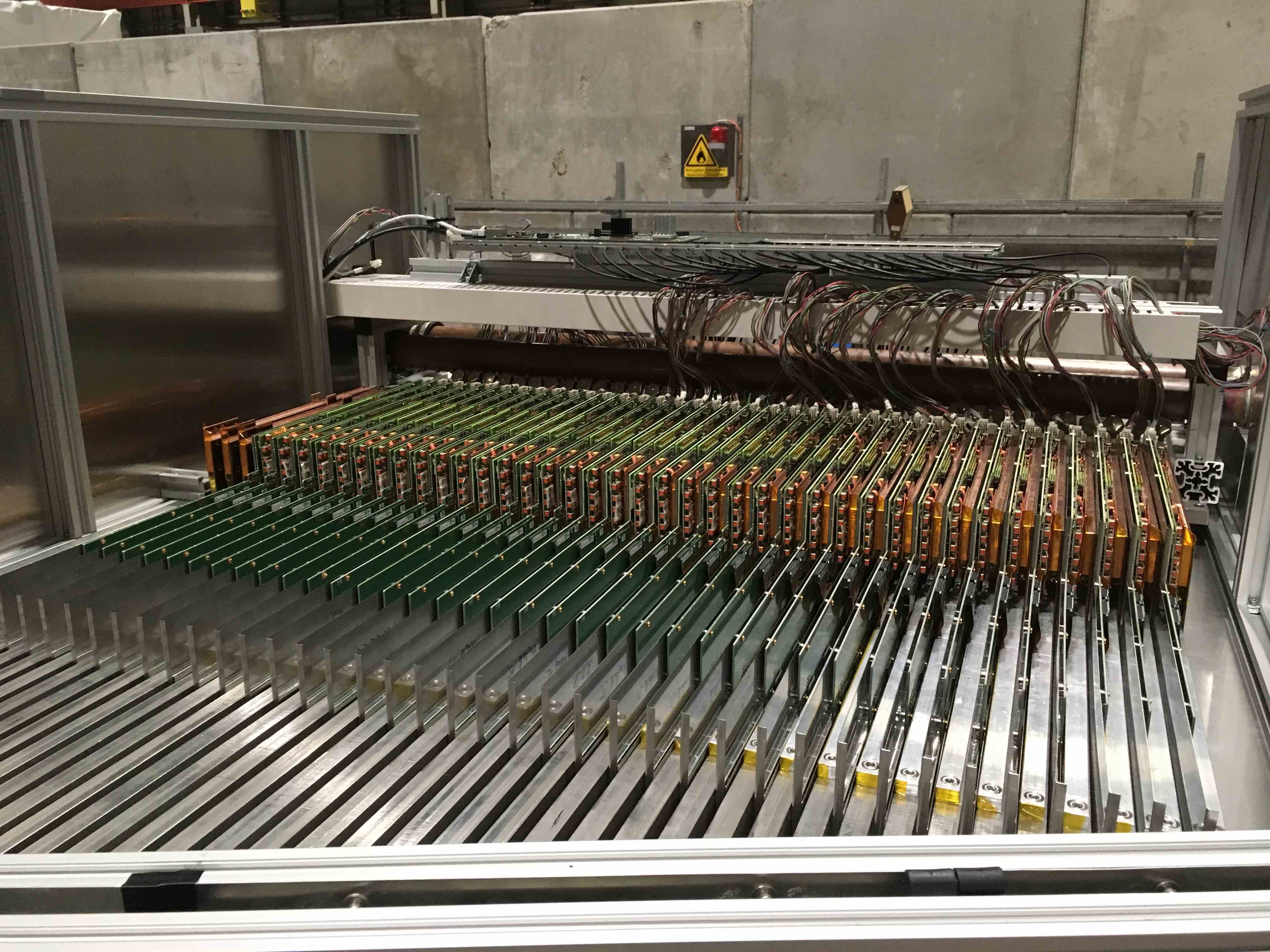}
\caption{SiPM-on-tile AHCAL engineering prototype.}
\label{fig:AbsorberStructure}
\end{figure}

\begin{figure}
\centering
\includegraphics[width=0.47\textwidth]{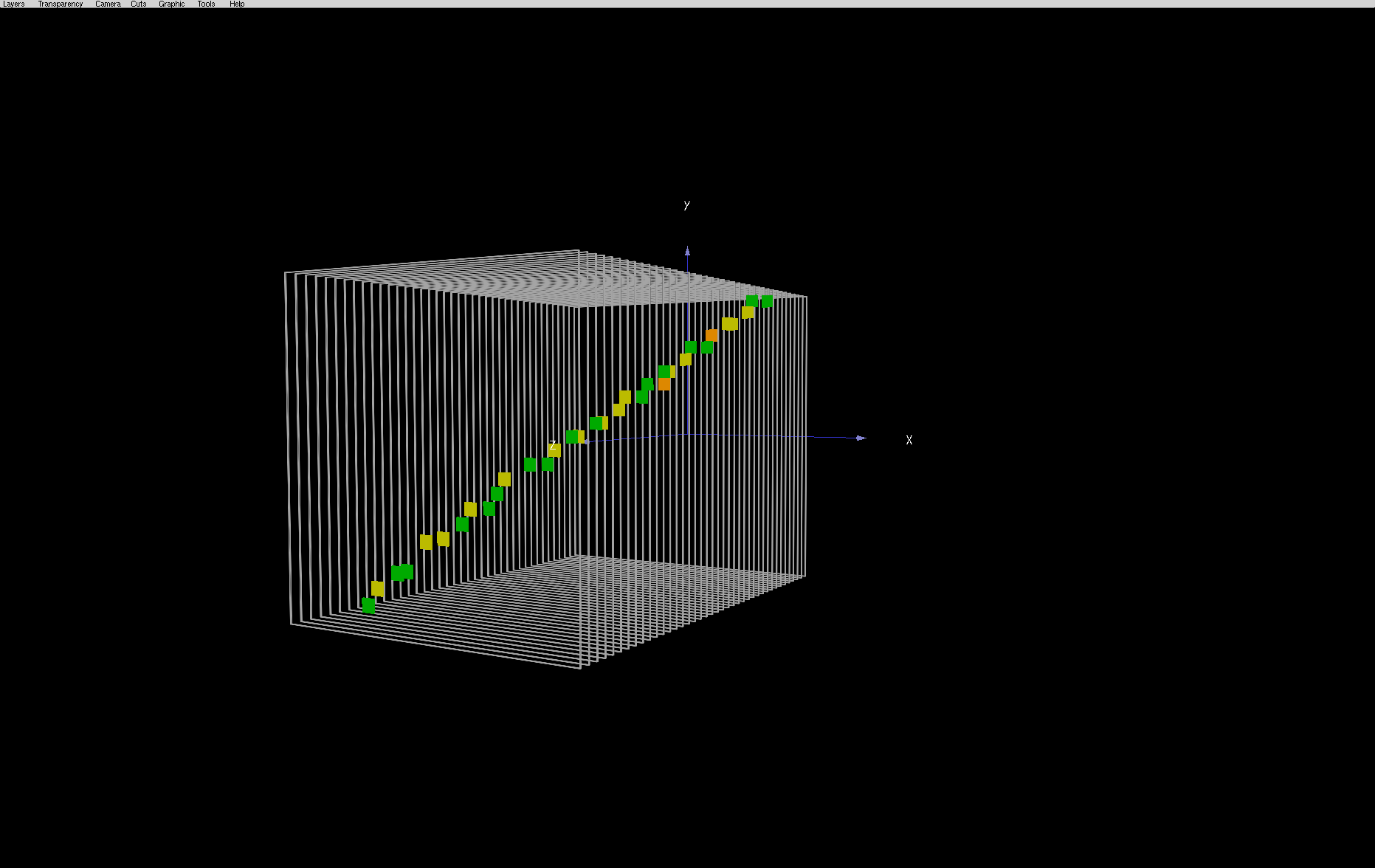}\hfill
\includegraphics[width=0.47\textwidth]{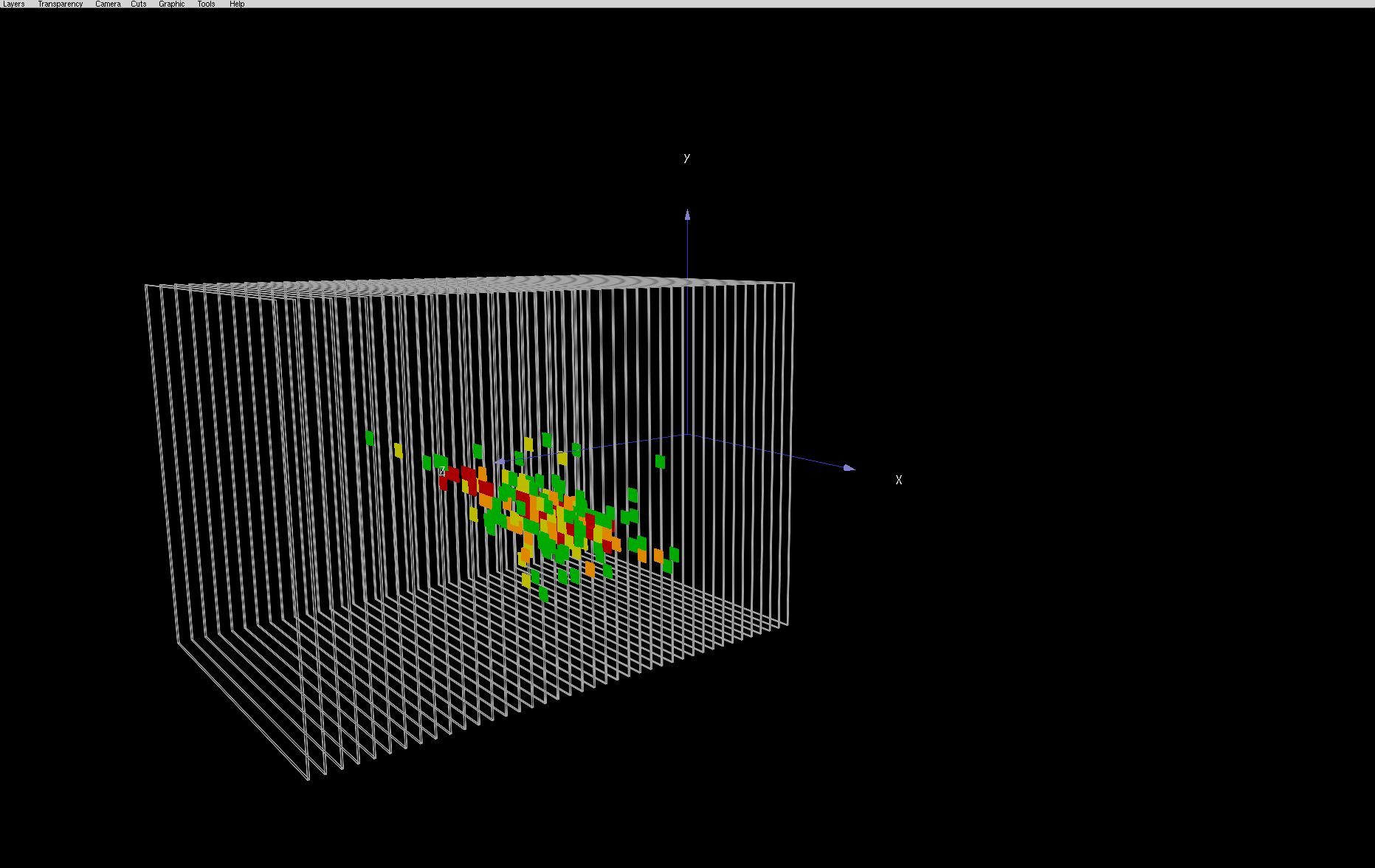}
\caption{Event display of
cosmic muons. Left: minimum-ionizing track. Right: muon-induced shower in the calorimeter volume.}
\label{fig:EventDisplay}
\end{figure}

Finally the calibrated layers were assembled into the absorber stack and connected to data concentration, power distribution and cooling services. Figure~\ref{fig:AbsorberStructure} shows the active layers with connected services inserted in the absorber structure.
The full prototype has been commissioned with cosmic muons, exploiting its self-triggering capabilities; see Figure~\ref{fig:EventDisplay}.
Two event displays are shown from cosmic rays interacting in the calorimeter. The top figures is a straight track from a minimum-ionising muon and the bottom is most likely a shower developed from an inelastic interaction of a muon with the absorber material.

\section{Beam test at the CERN SPS}

The prototype has been installed in the test beam for data taking at the CERN SPS. During two periods in May and in June 2018, several $10^7$ events with muon tracks, as well as electron and pion showers in the energy ranges  10 -- 100~GeV and 10 -- 200~GeV, respectively, have been recorded. The data taking rate averaged over the about 5~s long spills was up to 400 events per second.

\begin{figure}
\centering
\includegraphics[width=0.99\textwidth]{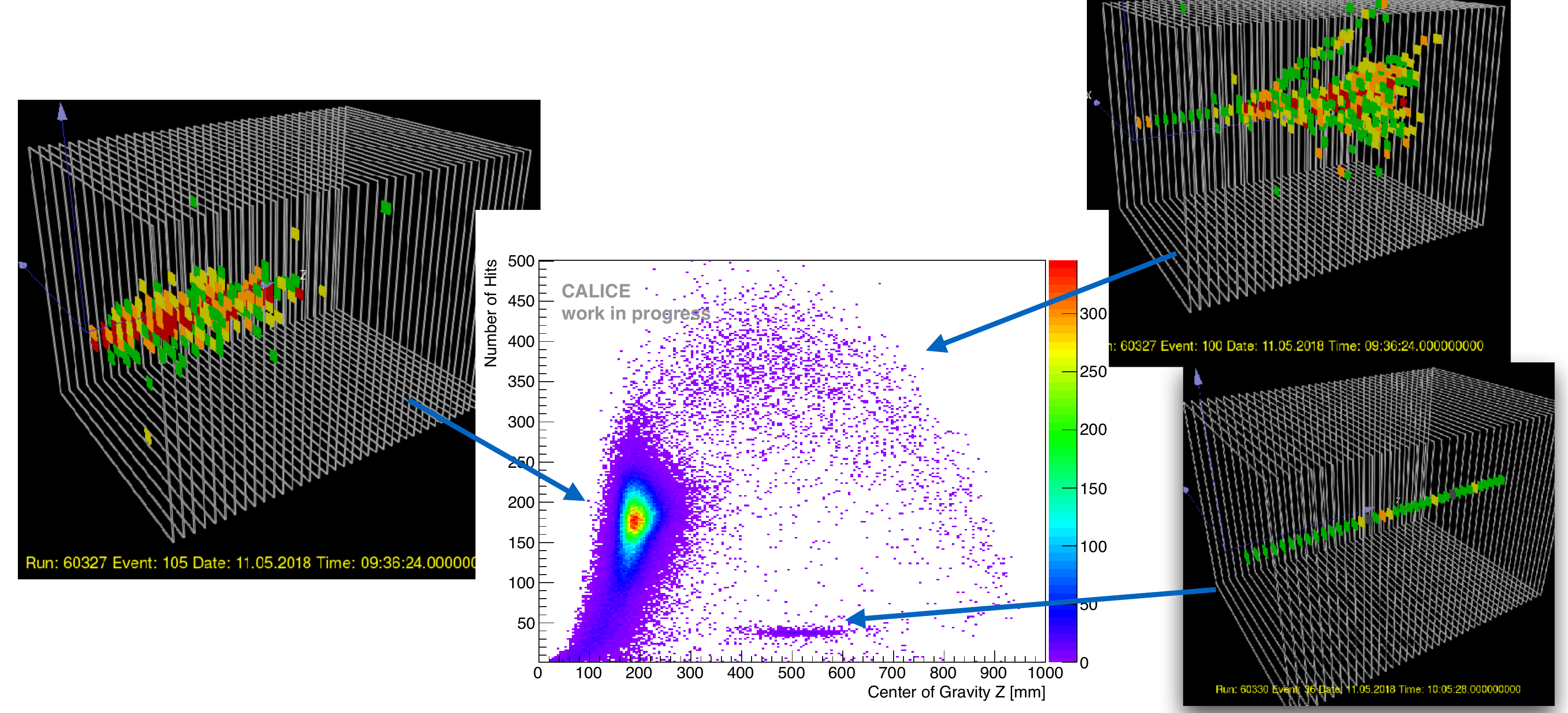}
\caption{Distribution of the number of hits vs.\ the hit-energy weighted centre-of-gravity (cog); event displays of typical
electron, hadron and muon events.} 
\label{fig:nhit-vs-z}
\end{figure}

Figure~\ref{fig:nhit-vs-z} from the quasi-instantaneous data quality monitoring shows the distribution of the number of hits vs.\ the hit-energy weighted centre-of-gravity (cog) along the beam axis z for an electron run with a beam momentum of 50~GeV/c and admixtures of muons and hadrons. The different particle types populate different regions of the plot, which is illustrated by the associated event displays. 

While electron showers are characterised by a relatively narrow distribution of number of hits and a cog near the front face of the detector, hadrons exhibit a wider distribution of the cog, and a larger number of hits, decreasing as the cog moves towards the rear of the detector, and leakage increases.
Muons appear as a narrow band with $\sim 38$ hits and a cog on z ay about half the depth of the detector. The figure shows how the detailed topological information of the AHCAL can be used for the identification of particle types. The width of the distribution for electrons, and tails towards lower number of hits, suggest a compromised beam quality for the May period shown here, which indeed was resolved for the June period.

\section{Conclusions}
A highly granular hadron calorimeter prototype with 21888 channels, based on 3$\times$3~cm$^2$ scintillator tiles and SiPMs integrated with the embedded read-out electronics, has been successfully constructed and operated in test beams. The scalable design and automated construction and quality assurance procedures validate the concept for linear collider detector applications. This has also inspired the design of the scintillator section of the CMS end-cap calorimeter upgrade for the high luminosity phase of the LHC~\cite{Collaboration:2293646}.
The rich data sample collected in the two test beam periods in 2018 will be used for shower separation studies based on 5-dimensional reconstruction algorithms exploiting the high spatial, energy and time resolution of this novel detector. 

\section*{Acknowledgments}
We would like to thank our CALICE colleagues for many enlightening discussions and continuous support. 
This project has received funding from the European UnionÕs Horizon 2020 Research and Innovation programme under Grant Agreement no. 654168.

\section*{References}
\bibliography{CALICE}

\end{document}